\documentclass{article}
\usepackage{arxiv}

\usepackage[utf8]{inputenc} 
\usepackage[T1]{fontenc}    
\usepackage{hyperref}       
\usepackage{url}            
\usepackage{booktabs}       
\usepackage{amsfonts}       
\usepackage{nicefrac}       
\usepackage{microtype}      
\usepackage{lipsum}
\usepackage[version=3]{mhchem} 
\usepackage{mathptmx}
\usepackage{multirow}
\usepackage{hyperref}
\usepackage{braket}
\usepackage{rotating}
\usepackage{color}
\usepackage{graphicx}
\usepackage{amsmath}
\usepackage[square,numbers,compress]{natbib}
\usepackage{notes2bib}

\newcommand{\fig}{Fig.}

\newcommand{\figref}[1]{\fig~\ref{#1}}

\newcommand{\Tabref}[1]{Table~\ref{#1}}

\renewcommand{\eqref}[1]{Eq.~(\ref{#1})}

\newcommand{\secref}[1]{Section~\ref{#1}}

\title{Gaussian Moments as Physically Inspired Molecular Descriptors for Accurate and Scalable Machine Learning Potentials}

\author{    Viktor Zaverkin\\
              Institute for Theoretical Chemistry\\
              University of Stuttgart\\
              Pfaffenwaldring 55,\\
              70569 Stuttgart, Germany \\
              \And
              Johannes Kästner$^1$\\
              Institute for Theoretical Chemistry\\
              University of Stuttgart\\
              Pfaffenwaldring 55,\\
              70569 Stuttgart, Germany \\}

\begin{document}
\footnotetext[1]{E-Mail: \texttt{kaestner@theochem.uni-stuttgart.de}}
\maketitle

\begin{abstract}
Machine learning techniques allow a direct mapping of atomic positions and nuclear charges to the potential energy surface with almost ab-initio accuracy and the computational efficiency of empirical potentials. In this work we propose a machine learning method  for constructing high-dimensional potential energy surfaces based on feed-forward neural networks. As input to the neural network we propose an extendable invariant local molecular descriptor constructed from geometric moments. Their formulation via pairwise distance vectors and tensor contractions allows a very efficient implementation on graphical processing units (GPUs). The atomic species is encoded in the molecular descriptor, which allows the restriction to one neural network for the training of all atomic species in the data set. We demonstrate that the accuracy of the developed approach in representing both chemical and configurational spaces is comparable to the one of several established machine learning models. Due to its high accuracy and efficiency, the proposed machine-learned potentials can be used for any further tasks, for example the optimization of molecular geometries, the calculation of rate constants or molecular dynamics.
\end{abstract}


\section{Introduction}

Most applications in computational chemistry require the use of potential energy surfaces (PES). The PES is a multidimensional real-valued function of atomic coordinates. It can be obtained by the solution of the electronic Schrödinger equation in the Born--Oppenheimer approximation~\cite{Born27}. For the estimation of individual points on the PES different 
techniques can be used, from ab initio electronic structure theory to empirical fits by force fields.
Especially highly accurate estimates are computationally expensive, thus applications that require energies and forces for a large number of atomic configurations, like molecular dynamics (MD) or geometry optimization, require significant amounts of computational time.

MD simulations of big systems, e.g. proteins or other macromolecules, are currently infeasible at the ab-initio level of theory. In such cases empirical force fields provide the necessary computational efficiency at the drawback of limited transferability~\cite{Mackerell04} and their general inability to describe bond-formation and bond-breaking. Therefore, a method which allows a direct mapping of atomic positions and nuclear charges to the PES, i.e.~$f: \{z_i, \mathbf{r}_i\} \mapsto E$, with maximal accuracy is required. 

Machine learning (ML) techniques can be applied for an efficient approximation of the PES, since, once trained, they hold the promise to combine the accuracy of ab-initio electronic structure methods with the efficiency of empirical force fields. For chemical applications several ML techniques can be used to predict a variety of chemical and physical properties of molecules and solids. The most frequently used approach is feed-forward neural networks (NN).

The construction of a reliable machine-learned mapping from atomic positions to potential energies requires a carefully chosen representation of the input to the ML algorithm defined by the atomic coordinates and nuclear charges. This is because the ML methodology doesn't exploit any information about the physics of the problem, in our case neither the invariance of a chemical system with respect to translation, reflection, rotation of the whole molecule nor to permutation of atoms with the same nuclear charge (atomic species). Therefore, a transformation to a suitable set of coordinates, i.e. a suitable descriptor, is required in order to obtain the desired accuracy in energy and gradient predictions. 

Several descriptors for ML models have been proposed. Some of the approaches split the molecules into atomic contributions and use hand-crafted descriptors, e.g. atom-centered symmetry functions (ACSF)~\cite{Behler07, Behler10}, power spectra or bispectra of spherical harmonics~\cite{Bartok10, Bartok13, Khorshidi16, Unke18, Kocer19}, or geometric moments~\cite{Shap16, Gub18}. Others use the Coulomb matrix of the whole molecule~\cite{Rupp12}. A different class of models is referred to as message-passing high-dimensional NNs, which learn to construct invariant features in a data-driven manner~\cite{Unke19, Schuett17_2, Schuett17, Schuett18, Lubb18, Schuett19}. Most of the methods based on hand-crafted descriptors are limited to only a few atomic species~\cite{Behler07, Behler10, Artrith17, Yao18}, smaller systems~\cite{Chmiela17, Chmiela18}, or fail to approach the accuracy of $1$~kcal/mol with respect to the underlying ab-initio method~\cite{Linienfeld15} required for chemical applications.

The requirements of a PES fit for successful application in chemistry are summarized in the following. It has to approximate the PES sufficiently accurately with an error below 1 kcal/mol in the energies with respect to the underlying ab-initio method and a comparable error for the forces. The approximation should be differentiable with respect to the atomic coordinates to allow for the calculation of forces and Hessians. It has to fulfill the invariances mentioned above: translation, rotation, permutation of like atoms. The fit should also be systematically improvable, i.e.\ the accuracy of predictions should increase with increasing size of the training data set. Finally, the machine learning model should be general, i.e. it should be transferable between similar systems and their configurations~\cite{Behler11}. Unfortunately, existing models and respective potential energy surfaces fulfill only a subset of these requirements.

In this work, we introduce a novel, physically inspired molecular descriptor, which can be used as input for any ML algorithm. We refer to it as Gaussian Moments (GM) since it was inspired by Gaussian-type atomic orbitals and derived from geometric moments previously used for pattern recognition~\cite{Fluss9, Fluss16, Suk11, Yang14}. We have chosen feed-forward NNs as an ML method for our applications. In addition to the structural description, we encode the information about the atomic species in the molecular representation. This allows us to use a single NN for all atomic species, in contrast to using an individual NN for each species as frequently necessary previously~\cite{Behler07, Artrith17, Khorshidi16, Yao18, Zhang19}. 
It is shown that the ML potentials built with the GM descriptor match or improve upon the state-of-the-art performance on standard benchmark data sets.

This paper has the following structure: first we formulate the molecular representation based on GMs and explain our machine learning model describing details on its training. Then in \secref{sec:sec3} we apply our machine learning model to the QM9~\cite{Rudd12, Rama14}, MD17~\cite{Chmiela17, Schuett17_2, Chmiela18}, and ISO17~\cite{Rama14, Schuett17, Schuett17_2} benchmark data sets and compare it to various models published in the literature. Additionally, we use it to predict vibrational frequencies based on a newly generated training set. The concluding remarks are given in \secref{sec:sec4}.

\section{\label{sec:sec2}Method}

As mentioned above, a suitable
descriptor, which converts atomic coordinates into ML input, should ensure the same global invariances as the physical
system. These are (1) the global rotation, (2) the translation, and (3) the reflection
of a molecular structure, as well as (4) the exchange of atoms of the same atomic species, i.e. with the same nuclear charge~$Z$.
One simple solution, which satisfies the requirements (1)--(3), can be
constructed using just the scalar product of vectors~$\mathbf{r}_{ij}$ from the position of a central atom $i$ to the positions $j$ of all other atoms, resulting in the Weyl matrix~\cite{Weyl46} $\Sigma_i$ 
\begin{equation}
\label{eq:eq_weyl}
\Sigma_i = 
\left(\begin{array}{cccc}
\mathbf{r}_{i1}\cdot\mathbf{r}_{i1} & \mathbf{r}_{i1}\cdot\mathbf{r}_{i2} & \cdots \\
\mathbf{r}_{i2}\cdot\mathbf{r}_{i1} & \mathbf{r}_{i2}\cdot\mathbf{r}_{i2} & \cdots \\
\vdots & \vdots & \ddots
\end{array}\right).
\end{equation}

However, any molecular system is invariant with respect to the exchange of two atoms of the same type. Therefore, a proper molecular representation has to incorporate this property as well. Unfortunately, introducing the permutation invariance into the above representation makes it intractable whenever one deals with large systems and, moreover, can violate the differentiability of the molecular representation~\cite{Bartok13}. Therefore, the main focus of this section is to introduce a class of molecular representations, which satisfies permutation invariance and is at least a $C^2$ function of the atomic positions, i.e. it is at least twice differentiable. 

\subsection{\label{sec:sec2.1}Molecular Descriptor}


The methodology of this study is based on the fact that the PES is the expectation value of electronic Hamiltonian $\hat{H}$, i.e. it is a solution of electronic Schrödinger equation
\begin{equation}
\hat{H}\Psi\left(\mathbf{r}\right) = E\left(\mathbf{r}\right)\Psi\left(\mathbf{r}\right).
\end{equation}
Here $\Psi$ is the electronic wave function which depends on the atomic position vector $\mathbf{r}$. Thus, the energy of a molecular system is a functional of the electronic wave function
\begin{equation}
E = \mathcal{F}\left[\Psi\right].
\end{equation}
The electronic wave function can be efficiently expanded into atom-centered Gaussian-type orbitals, which inspired our choice of the molecular descriptors. Note that the descriptor uses exclusively atomic positions rather than electronic coordinates. 

\begin{figure}
\includegraphics[width=\linewidth]{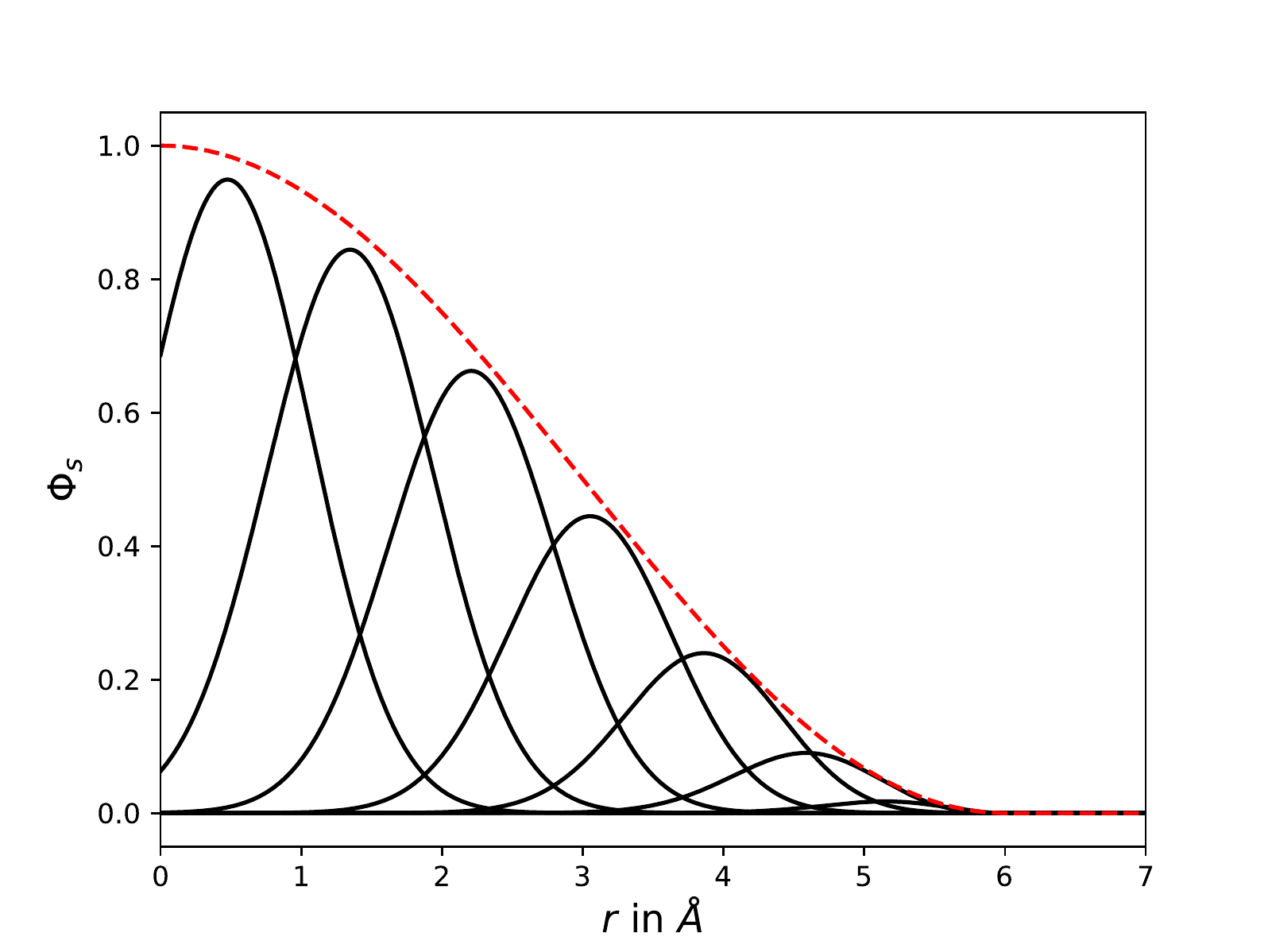}
\caption{Radial basis functions $\Phi_{s}(r)$ (black) and the cutoff function $f_\text{cut}(r)$ (red, dashed) for $R_\text{max} = 6.0$~{\AA} and $N_\text{Gauss}=7$.}
\label{fig:fig_radial}
\end{figure}

We split the descriptor for the whole chemical system into functions, which describe the environment of each atom individually. Those can subsequently be combined to describe whole molecular or periodic systems. The environment of each atom is described by a function reminiscent of a Gaussian-type orbital~\cite{Boys50} (GTO)
\begin{equation}
\label{eq:eq_gto}
\varphi_{s, l_x, l_y, l_z}\left(\mathbf{r}\right) = \frac{x^{l_x}y^{l_y}z^{l_z}}{r^L} \Phi_{s}\left(r\right)
\end{equation}
with $\mathbf{r}=\left(x, y, z\right)$ being an atom's position relative to a
central atom, $r$ being its absolute value, and $L$ defined as $L=l_x+l_y+l_z$. The pre-factor
$x^{l_x}y^{l_y}z^{l_z}/r^L$ covers the angular dependence of the GTO, which we will deal with
in \eqref{eq:eq_psiexample}. 
The radial part $\Phi_{s}\left(r\right)$ was chosen to be a single normalized Gaussian with a radial cutoff defined as
\begin{equation}
\Phi_{s}\left(r\right) = 
\left(\frac{2N_\text{Gauss}^2}{\pi R_\text{max}^2}\right)^{1/4}
e^{-\frac{N_\text{Gauss}^2}{R_\text{max}^2}\left( r - \gamma_s\right)^2}f_\text{cut}\left(r\right).
\end{equation}
The width of each Gaussian depends on the total number $N_\text{Gauss}$ of functions used and the cutoff radius $R_\text{max}$. Each Gaussian is centered at $\gamma_s$, which is chosen evenly spaced between $R_\text{min}$ and $R_\text{max}$,
\begin{equation}
\label{eq:eq_spacing}
\gamma_s=R_\text{min}+\frac{s-1}{N_\text{Gauss}-1} (R_\text{max}-R_\text{min})
\end{equation}
with $s$ being an index from 1 to $N_\text{Gauss}$. As discussed in \secref{sec:sec3}, we typically use $N_\text{Gauss}=7$ and $R_\text{min}=0.5$~\AA{}. $R_\text{max}$ depends on the specific case.  Note that $\gamma_s$ is defined for the whole data set. An example of the radial basis functions $\Phi_{s}\left(r\right)$ is shown in \figref{fig:fig_radial}.

Each radial function incorporates a cutoff function $f_\text{cut}(r)$, which restricts the descriptor to the local neighborhood of the atom and decays smoothly to zero at the cutoff radius $R_\text{max}$. In this work we have chosen the cosine cutoff function~\cite{Behler07}, see \figref{fig:fig_radial},
\begin{equation}
f_\text{cut}\left(r\right)=\left\{
\begin{array}{lr}
\frac{1}{2}\left(\cos\left(\pi \frac{r}{R_\text{max}}\right)+1\right) & r \leq R_\text{max},\\
0& r> R_\text{max}.
\end{array}\right.
\end{equation}
Periodic boundary conditions are incorporated by including the periodic images of atoms in the local neighborhood. The GM descriptor is constructed from the coordinates of the image atoms and the atoms within the cell. However, a more thorough discussion of periodic calculations is beyond the scope of this work.

In the next step we form a linear combination of the atomic ``wave'' functions $\varphi_{s, l_x, l_y, l_z}\left(\mathbf{r}\right)$, similar to the linear combination of atomic orbitals (LCAO), again inspired by quantum chemistry. The total molecular wave function centered at an atom $i$ reads
\begin{equation}
\label{eq:eq_psi}
\Psi_{i,L,s} = \sum_{j\ne i}^{N_\text{at}} \beta_{Z_i, Z_j,s} \varphi_{s, l_x, l_y, l_z}(\mathbf{r}_{ij}),
\end{equation}
where $Z_i$ and $Z_j$ are the nuclear charges of the central atom $i$ and its
atomic neighbors $j$. The coefficients $\beta_{Z_i, Z_j, s}$ distinguish
between nuclear charges and radial shells. They are optimized in the training
procedure. For a given $i$, $L$, and $s$, $\Psi_{i,L,s}$ is a tensor of rank~$L$.

\eqref{eq:eq_psi} preserves invariances (2) and (4) by construction. $\Psi_{i,L,s}$ is invariant with respect to translations (2) owing to its dependence on the atomic distance vectors $\mathbf{r}_{ij}$. The invariance with respect to permutation of like atoms (4) is ensured by the sum. However, the pre-factor $x^{l_x}y^{l_y}z^{l_z}/r^L$ still violates invariance with respect to rotation and reflection for $L>0$. Consequently, $\Psi_{i,L,s}$ cannot be used directly as input to ML algorithms and further treatment is necessary.

One can interpret $L$ as an angular momentum similar to spherical
harmonics. For example, $L=0$ corresponds to the shape of a spherically symmetric $s$-orbital,
$L=1$ corresponds to the shape of a $p$-orbital, $L=2$ to that of a $d$-orbital, and so on.
To construct a rotationally invariant basis, we look deeper into the mathematical properties of $\Psi_{i,L,s}$. GTO functions in \eqref{eq:eq_gto} can be written as a Cartesian tensor. For $L={0,1,2}$ we can write $\Psi_{i,L,s}$, when rewriting the angular dependence in terms of atomic distance vectors $\mathbf{r}_{ij}$ rather than in terms of its components, as
\begin{equation}
\label{eq:eq_psiexample}
\begin{split}
\Psi_{i,0,s} = \sum_{j\neq i}^{N_\text{at}} \beta_{Z_i, Z_j, s}\Phi_{s}\left(r_{ij}\right),\\
\boldsymbol{\Psi}_{i,1,s} = \sum_{j\neq i}^{N_\text{at}} \beta_{Z_i, Z_j, s}\frac{\mathbf{r}_{ij}}{r}\Phi_{s}\left(r_{ij}\right), \\
\boldsymbol{\Psi}_{i,2,s} = \sum_{j\neq i}^{N_\text{at}} \beta_{Z_i, Z_j, s}\frac{\mathbf{r}_{ij}\otimes\mathbf{r}_{ij}}{r^2}\Phi_{s}\left(r_{ij}\right),
\end{split}
\end{equation}
where $\otimes$ denotes the tensor product. For an arbitrary angular momentum $L$ one can write
\begin{equation}
\label{eq:eq_psifinal}
\boldsymbol{\Psi}_{i,L,s} = \sum_{j\neq i}^{N_\text{at}} \beta_{Z_i, Z_j, s}
\underbrace{\mathbf{r}_{ij}\otimes\cdots\otimes\mathbf{r}_{ij}}_{L\text{ times}}\frac{1}{r^L}\Phi_{s}\left(r_{ij}\right).
\end{equation}
The tensor $\mathbf{r}_{ij}\otimes\cdots\otimes\mathbf{r}_{ij}$ has rank $L$ and will, in the following discussion, be referred to as $T_{i_1,i_2,\dots,i_L}=\left(\mathbf{r}_{ij}\otimes\cdots\otimes\mathbf{r}_{ij}\right)_{i_1,i_2,\ldots,i_L}$ to simplify the notation. Since $T_{i_1,i_2,\dots,i_L}$ is a Cartesian tensor it behaves under rotation according to the rule
\begin{equation}
\label{eq:eq10}
\hat{T}_{\alpha_1, \alpha_2,\dots,\alpha_L} = R_{\alpha_1,i_1}R_{\alpha_2,i_2}\cdots R_{\alpha_L,i_L}T_{i_1,i_2,\dots,i_L},
\end{equation}
where $R_{\alpha_1,i_1}$ is an arbitrary orthonormal matrix, e.g. a rotation or reflection. From linear algebra it is known that any full contraction of a Cartesian tensor or of a product of Cartesian tensors is a rotationally invariant scalar. The radial function doesn't affect this property due to its inherent invariance with respect to rotations. The same holds for reflections. Consequently, an invariant basis, which satisfies all the requirements can be constructed by calculating the full contractions of the molecular wave function.

Note that the concept of constructing rotational invariants using contractions of Cartesian tensors was initially introduced by J. Flusser, T. Suk et. al.~\cite{Fluss9, Fluss16, Suk11, Yang14}, where geometric and Gaussian--Hermite moments were used to address pattern recognition problems. Additionally, geometric moments were used to construct rotationally invariant bases for linear regression in PES construction~\cite{Shap16, Gub18}.

\begin{figure}
\begin{center}
\includegraphics[width=\linewidth]{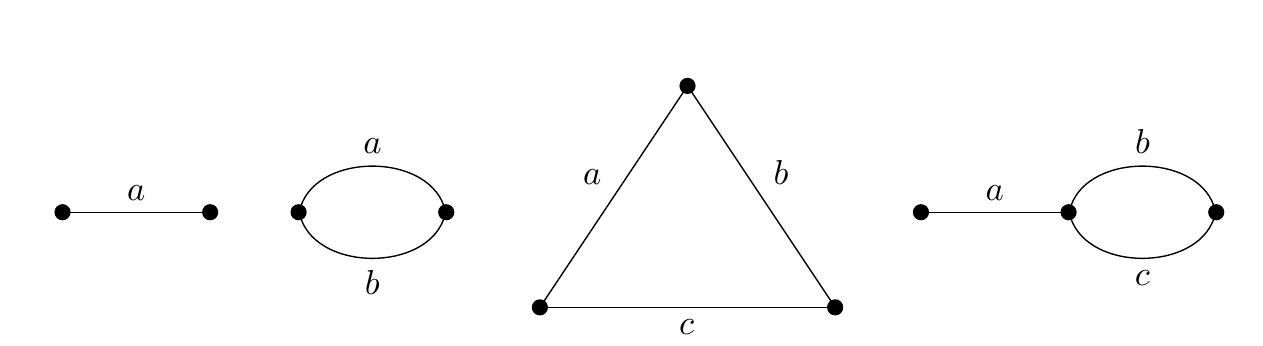}
\end{center}
\caption{Generating graphs for the tensor contractions (12.2), (12.3), (12.6), and (12.7) of \eqref{eq:eq_contractions}.}
\label{fig:fig_graphs}
\end{figure}

Inspired by previous work on invariants obtained using geometric moments~\cite{Fluss9, Fluss16, Suk11, Yang14, Shap16, Gub18}, we will refer to scalars obtained by contracting $\Psi_{i,L,s}$ as Gaussian Moments (GM). To simplify the generation of contractions we employed graphs~\cite{Suk11}. Some examples are shown in \figref{fig:fig_graphs}. However, one can find a direct correspondence to index-matrices~\cite{Shap16} and use them instead. 

In general for the representation of a molecular structure at least a $\left(3N_\text{at}-6\right)$-dimensional descriptor is needed. This can be fulfilled by using only rather few contractions. It turned out to be sufficient to restrict the total angular momentum to $L \leq 3$ and the maximal number of contracted tensors to $3$.  This results in a total of eight
contractions, i.e. Gaussian moments, that we used throughout this work:
\begin{equation}
\label{eq:eq_contractions}
\begin{split}
\rho_{i, s_1} = \boldsymbol{\Psi}_{i,0,s_1},\\
\rho_{i, s_1, s_2} = \left(\boldsymbol{\Psi}_{i,1,s_1}\right)_a \left(\boldsymbol{\Psi}_{i,1,s_2}\right)_a,\\
\rho_{i, s_1, s_2} = \left(\boldsymbol{\Psi}_{i,2,s_1}\right)_{a,b}\left(\boldsymbol{\Psi}_{i,2,s_2}\right)_{a,b},\\
\rho_{i, s_1, s_2} = \left(\boldsymbol{\Psi}_{i,3,s_1}\right)_{a,b,c}\left(\boldsymbol{\Psi}_{i,3,s_2}\right)_{a,b,c},\\
\rho_{i, s_1, s_2, s_3} = \left(\boldsymbol{\Psi}_{i,2,s_1}\right)_{a,b}  \left(\boldsymbol{\Psi}_{i,1,s_2}\right)_{a}  \left(\boldsymbol{\Psi}_{i,1,s_3}\right)_{b},\\
\rho_{i, s_1, s_2, s_3} = \left(\boldsymbol{\Psi}_{i,2,s_1}\right)_{a,b}  \left(\boldsymbol{\Psi}_{i,2,s_2}\right)_{a,c}  \left(\boldsymbol{\Psi}_{i,2,s_3}\right)_{b,c},\\
\rho_{i, s_1, s_2, s_3} = \left(\boldsymbol{\Psi}_{i,1,s_1}\right)_{a}  \left(\boldsymbol{\Psi}_{i,3,s_2}\right)_{a,b,c}  \left(\boldsymbol{\Psi}_{i,2,s_3}\right)_{b,c},\\
\rho_{i, s_1, s_2, s_3} = \left(\boldsymbol{\Psi}_{i,3,s_1}\right)_{a,b,c}  \left(\boldsymbol{\Psi}_{i,3,s_2}\right)_{a,b,d}  \left(\boldsymbol{\Psi}_{i,2,s_3}\right)_{c,d}.
\end{split}
\end{equation}
Here, Einstein's notation was used for tensor contractions, i.e. the sum is taken over double indices, to simplify the expressions. All these tensors are symmetric. We use only upper triangular entries as descriptors.

In total using $N_\text{Gauss}=7$ and all contractions given in \eqref{eq:eq_contractions} we obtained $7+28\cdot 3+84\cdot 4= 427$ rotationally invariant scalars for each atom. These constitute the molecular descriptor, which was used as input for the NN in \secref{sec:sec3}. All elements of the molecular descriptor depend on the atomic species of the central atom and its atomic neighborhood. This dependence in encoded using the coefficients $\beta_{Z_i, Z_j, s}$ which are optimized during training.

Contractions of two wave functions can be related to electronic densities with an angular momentum $L$. 
Electronic densities were recently used for the construction of a molecular representation in ML~\cite{Zhang19}. However, the approach presented here is more general than electronic densities as it allows to contract more (and less) than two wave functions to construct rotational invariants. Thus, much more insight in the angular and radial distribution of the atomic environment can be incorporated into the machine learning algorithms at the same computational cost.

\subsection{\label{sec:sec2.2}Atomistic Neural Networks}

Artificial neural networks (NN) have been proven to be capable of approximating any non-linear functional relationship~\cite{Hornik91}. Therefore, they are of particular interest for reproducing high-dimensional potential energy surfaces (PES). 
Behler and Parrinello suggested a construction, which allows the application of NNs to systems of different sizes~\citep{Behler07}. In their approach the total energy $\hat{E}$ of a molecular system is decomposed into a sum of atomic contributions~$\hat{E}_i$
\begin{equation}
\label{eq:eq_esum}
\hat{E} = \sum_{i=1}^{N_\text{at}}\hat{E}_i = \sum_{i=1}^{N_\text{at}}\text{NN}_{Z_i}\left(\mathbf{x}_\text{in}^{(i)}\right),
\end{equation}
where $\text{NN}_{Z_i}$ denotes the neural network output, and $\mathbf{x}_\text{in}$ is a molecular representation. In their approach an individual neural network $\text{NN}_{Z_i}$ is constructed and trained for each atomic species~$Z_i$. In our approach a similar construction is used. Since the Gaussian moment representation $\rho_{i, s_1, s_2,\dots}$ contains the information about the atomic species via the coefficients $\beta_{Z_i, Z_j, s}$, a single NN is constructed and trained for all species. This results in the expression for the total energy
\begin{equation}
\label{eq:eq16}
\hat{E} = \sum_{i=1}^{N_\text{at}}\text{NN}\left(\mathbf{x}_\text{in}^{(i)}=\{\rho_{i, s_1},\ \rho_{i, s_1, s_2},\ \dots\}\right).
\end{equation}
The approach presented here is atom-centered and, thus, allows the modeling of molecular systems with a variable number of atoms. 

In this work, a feed-forward neural network is used. In a feed-forward NN an input layer is connected to an output layer via one or multiple hidden layers. The information in the network passes only in a single direction towards the output layer. The local molecular descriptor, i.e.~$\mathbf{x}_\text{in}^{(i)}=\{
\rho_{i, s_1},\ \rho_{i, s_1, s_2},\ \dots\}$, provides the values of the neurons in the input layer, while the output of the NN is the atomic energy, $\hat{E}_i$. A linear transformation is applied to the input data for each layer followed by a non-linear activation function, i.e.~for two hidden layers
\begin{equation}
\label{eq:eq_network}
\mathbf{y}_\text{out} = \phi_\text{out}\Big( \phi_2\big( \phi_1\left(\mathbf{x}_\text{in}\mathbf{W}_1 + \mathbf{b}_1\right)\mathbf{W}_2 + \mathbf{b}_2\big)\mathbf{W}_\text{out} + \mathbf{b}_\text{out}   \Big),
\end{equation}
where $\mathbf{W}_k$ are the weight matrices, $\mathbf{b}_k$ are the biases, and $\phi_k$ are activation functions. For the output layer a linear activation function is used, whereas for the hidden layers non-linear activation functions are applied. In this work, a ``rectifier''-like function, the soft-plus function $\phi_i\left(x\right)=\ln\left(1+\exp\left(x\right)\right)$, was chosen as the non-linear activation function. We found it to perform better than other standard activation functions for the data sets used here.
In order to maximize the use of the non-linear region of the activation functions, the atomic energy is scaled and shifted as $\hat{E}_i = \sigma_{Z_i} y_{\text{out},i} + \mu_{Z_i}$. The parameters $\sigma_{Z_i}$ and $\mu_{Z_i}$ depend on atomic species and are optimized during the training procedure. The initialization of $\sigma_{Z_i}$ and $\mu_{Z_i}$ is performed by using the standard deviation and mean of the per-atom average of the reference energies in the training set to improve the convergence of the model. 

\begin{figure}
\includegraphics[width=\linewidth]{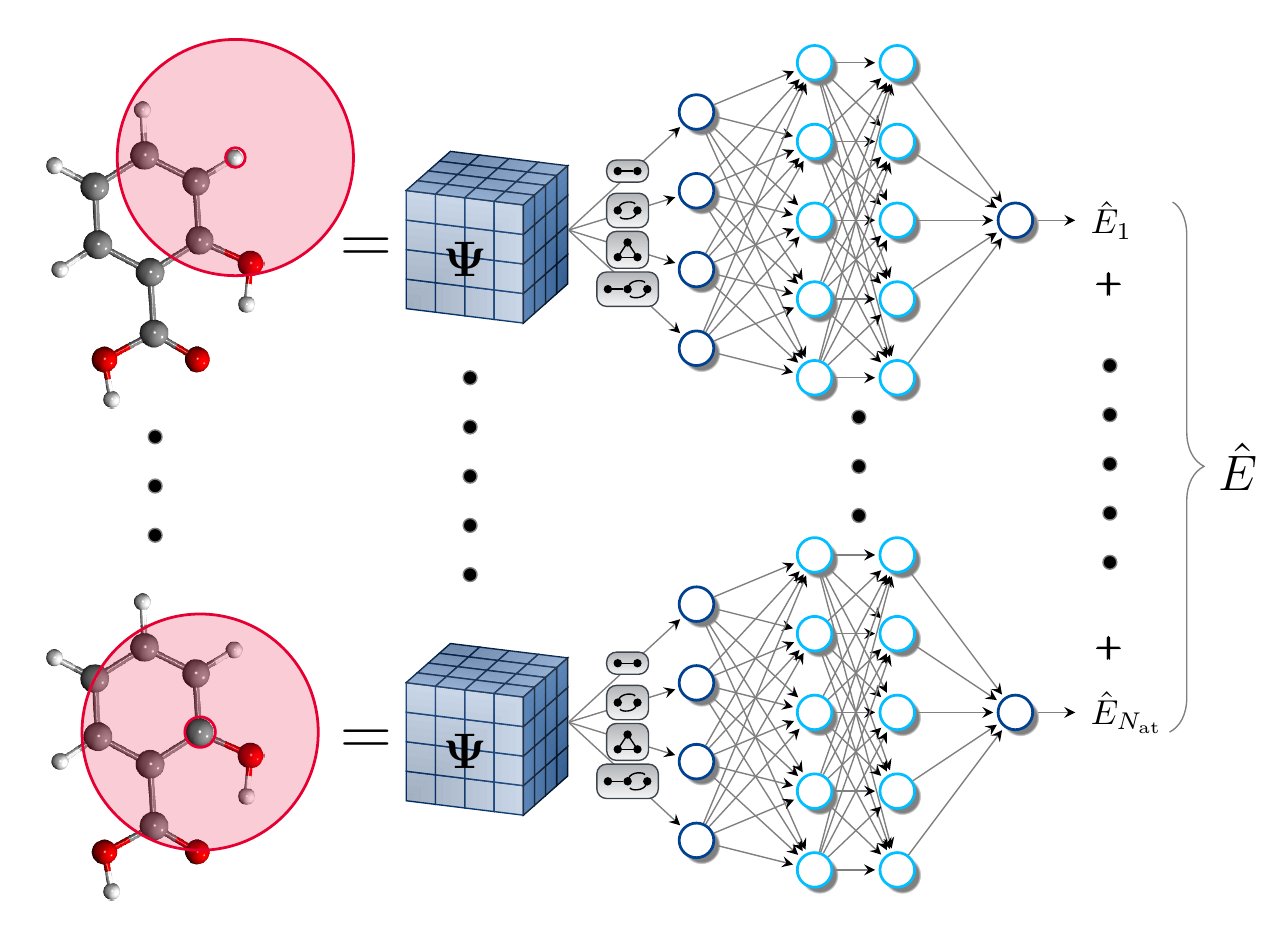}
\caption{Schematic representation of the model used in this work for calculating molecular energies and forces.}
\label{fig:fig_nnet}
\end{figure}

A schematic representation of an atom-centered feed-forward NN and the computational procedure of the presented GM-model is shown in \figref{fig:fig_nnet}.
First, a neighborhood of all atoms within the cutoff radius $R_\text{max}$ is assigned to each atom~$i$. Next, given the parameters $\gamma_s$, the radial functions $\Phi_{s}\left(r\right)$ are evaluated. Using the coefficients $\beta_{Z_i, Z_j, s}$, which are initiated randomly, the tensor-valued function $\boldsymbol{\Psi}_i$ centered at the atom $i$ is constructed. Then, the predefined tensor contractions are applied and the molecular representation $\rho$ is calculated. It is used as input to the feed-forward NN which outputs scaled atomic energies, $y_{\text{out},i}$. These are transformed back to non-scaled values, $\hat{E}_i$, which are summed up to result in the total energy of the system.

In total, two network architectures, a shallow and a deep NN, are constructed to test our model on benchmark data sets in \secref{sec:sec3}. The shallow network has two hidden layers with $\left[256, 128\right]$ nodes, respectively. The deep network consists of five hidden layers with $\left[1024, 512, 256, 128, 64\right]$ nodes each. We will refer to the shallow model as GM-sNN and to the deep model as GM-dNN.

\subsection{\label{sec:sec2.3}Training}

In this work, we are interested in the prediction of energies and forces and possibly Hessians in the future. Therefore, prior to describing the training procedure a few sentences are dedicated to the importance of the incorporation of forces into the training. For quantum chemical training data, obtaining forces for all atoms is about as computationally expensive as obtaining the energy. Thus, forces provide additional training data which are comparably cheap to obtain. Therefore, they are included in the training of the model.

To optimize weights and biases of each layer of the GM-model the training loss function is defined as
\begin{equation}
\label{eq:eq_loss}
\mathcal{L} = w_\text{E}\left|\left| \hat{E} - E^\text{ref}\right|\right|^2 + \frac{w_\text{F}}{3N_\text{at}}\sum_{i=1}^{N_\text{at}}\sum_{k=1}^3 \left|\left| \hat{F}_{i,k} - F^\text{ref}_{i,k}\right|\right|^2.
\end{equation}
To control the energy and force contribution during the training we define the adjustable parameters $w_\text{E}$ and $w_\text{F}$. The parameters were set to $w_\text{E}=1$ and $w_\text{F}=100$~\AA$^2${} for all models. The higher weight of the force error is motivated by the fact that forces alone determine the dynamics of a chemical system. Consequently, the accurate force prediction is most important for MD simulations. In case the model is trained only on energies the parameter $w_\text{F}$ is set to zero. The parameters, $w_\text{E}$ and $w_\text{F}$, were chosen according to performance tests of the GM-NN model. However, optimal values are likely to  depend on the system under study and the parameters should be adjusted accordingly. A more thorough investigation of the dependence of the performance on the parameters is planned for the future works.

The reference values for the force and energy are denoted by $E^\text{ref}$ and $\mathbf{F}^\text{ref}$, respectively. Atomic forces $\hat{\mathbf{F}}$ are calculated from the total energy $\hat{E}$ analytically by taking the partial derivative with respect to atomic positions. For an atom $i$ along the component $k \in \{x, y, z\}$ the atomic force is defined as
\begin{equation}
\label{eq:eq_force}
\begin{split}
\hat{F}_{i,k} \left(Z_1, Z_2, \dots, Z_{N_\text{at}}, \mathbf{r}_1, \mathbf{r}_2, \dots, \mathbf{r}_{N_\text{at}}\right) =\\ -\frac{\partial \hat{E}}{\partial r_{i,k}}\left(Z_1, Z_2, \dots, Z_{N_\text{at}}, \mathbf{r}_1, \mathbf{r}_2, \dots, \mathbf{r}_{N_\text{at}}\right).
\end{split}
\end{equation}

All models used in \secref{sec:sec3} were implemented in the Tensorflow~\cite{TF15} framework. Atomic forces were calculated using automatic differentiation~\cite{Baydin18} The training loss in \eqref{eq:eq_loss} was minimized using the AMSGrad optimizer~\cite{Reddi18} with $32$ molecules per mini-batch with an exception of the models trained on the ISO17~\cite{Rama14, Schuett17, Schuett17_2} data set, where a mini-batch of $128$ molecules was used. 
The learning rate was set to $10^{-3}$ for all models and kept constant throughout the whole training procedure. Each optimization took $5000$ training epochs with an exception of the models trained on $1000$ MD17~\cite{Chmiela17, Schuett17_2, Chmiela18} samples, where we optimized for 10,000 epochs.
Overfitting was prevented using the early stopping technique~\cite{Prechelt12}. After each epoch the training loss was evaluated on a validation set. After training, the model that performed best on the validation set was selected for further application on the test sets. So, although the validation data was not used directly in the training procedure, it indirectly influenced models chosen at the end.

\subsection{\label{sec:sec2.4}Scalability and Computational Cost}

To achieve linear scaling of the computational cost and memory usage, the GM-NN model uses atom neighbor lists as implemented in ASE~\cite{ase17}. This allows the calculation of the energy and gradient for a structure with up to $100,000$ atoms in less than 230 s on a single Intel Xeon CPU E5-2670 0. The memory required for the respective calculations with up to 25,000 atoms is about $9.7$ GB. This allows efficient training and inference on typical GPUs for large systems. Further information on the computational cost and memory usage, including details on the trained model, is provided in the Supporting Information.

\section{\label{sec:sec3}Results}

Here, we apply the NN model based on Gaussian Moments (GM-NN) to three well-established quantum chemistry data sets: QM9~\cite{Rudd12, Rama14}, MD17~\cite{Chmiela17, Schuett17_2, Chmiela18}, and ISO17~\cite{Rama14, Schuett17, Schuett17_2}. These data sets are designed such that different aspects of chemical space are covered. For all data sets, we report the mean absolute error (MAE) and the root mean square error (RMSE) in kcal/mol for the energies and in kcal/mol/\AA{} for the forces. 

The deep network model GM-dNN was tested only on large training sets, i.e.\ 50,000 training samples from the MD17 data set and 400,000 training samples from the ISO17 data set. The reason for this is that for smaller training sets, e.g.\ 1000 samples from the MD17 data set and the QM9 data set, the shallow GM-sNN model is already sufficient to reach an acceptable accuracy within the given number of training epochs. The deep architecture is prone to overfitting, especially for small training sets. The deep architecture is promising for large and complex training sets, because it is known that the additional hidden layers enhance the capability of neural networks to capture complexity and high non-linearity of functional dependence~\cite{Bengio07, Krizh12}.

The input layer for both architectures has $427$ neurons as discussed in \secref{sec:sec2.1}. The only remaining adjustable parameter of the descriptor is the cutoff radius $R_\text{max}$. It was set to $3.0$~{\AA} for the QM9 data set and to $4.0$~{\AA} for the MD17 and ISO17 data sets. In each experiment, the data set is split into a training set of size $N$ and a validation set containing $2000$ structures used for early stopping. The remaining data was used for testing the models.

\subsection{\label{sec:sec3.1} QM9}

QM9~\cite{Rudd12, Rama14} is a widely used benchmark for the prediction of several properties of molecules in equilibrium. Thus, all forces vanish. They were not included into the training loss function. Only shallow GM-sNN models were trained on the QM9 data set. 

The QM9 data set consists of 133,885 neutral, closed-shell organic molecules with up to 9 heavy atoms (C, O, N, F) and a varying number of hydrogen (H) atoms. The largest structure in the data set contains $29$ atoms in total. Since $3054$ molecules from the original QM9 data set failed a consistency test~\cite{Rama14}, we used only the remaining $130,831$ structures in the following experiments.

For QM9 a cutoff radius of $R_\text{max} = 3.0$ \AA{} was chosen. This is rather small compared to the $10$~\AA{} used in the message-passing architectures, e.g., SchNet~\cite{Schuett17, Schuett18} or PhysNet~\cite{Unke19}. However, the sphere defined by the small cutoff radius of $3.0$~{\AA} includes already a maximum of $24$ neighbors out of $28$ possible neighboring atoms for the largest structures in the data set. This holds for central atoms of the respective structures. For the side atoms smaller local environments can be found which can be transferred to the smaller structures in the data set. So, the smaller cutoff improves the ability of the model to generalize. Thus, the cutoff radius has to be increased only in the case some important interactions are neglected, which is not the case for the QM9 data set.

\begin{sidewaystable*}[htbp]
\centering
\caption{MAEs in kcal/mol for the energy prediction on the QM9 data set\cite{Rudd12, Rama14} for various models reported in the literature and different sizes of the training set. Results of the GM-sNN model are averaged over three and five independent randomly chosen training data sets, see text.}
\label{tab:table_qm9}
\begin{tabular}{cccccccc}
\hline
Training set size & DTNN~\cite{Schuett17_2} & SchNet~\cite{Schuett17,Schuett18} & PhysNet~\cite{Unke19} & HIP-NN~\cite{Lubb18} & MTM$_{16-28}$~\cite{Gub18} & Ref.~\citenum{Unke18} & GM-sNN\\
\hline
$1000$& $\cdots$&$\cdots$&$\cdots$&$\cdots$&$\mathbf{1.8}$&$1.85$&$2.16$\\
$5000$& $\cdots$&$\cdots$&$\cdots$&$\cdots$&$\mathbf{0.90}$\textsuperscript{\emph{b}}&$0.95$&$0.95$\\
$10,000$& $\cdots$&$1.28$\textsuperscript{\emph{a}}&$\cdots$&$\cdots$&$0.86$&$0.73$&$\mathbf{0.71}$\\
$25,000$& $1.04$&$0.80$\textsuperscript{\emph{a}}&$\cdots$&$\cdots$&$0.63$&$0.55$&$\mathbf{0.47}$\\
$50,000$& $0.94$ &$0.59$&$\mathbf{0.30}$&$0.35$&$0.41$&$0.46$&$0.36$\\
$100,000$& $0.84$&$0.34$&$\mathbf{0.19}$&$0.26$&$\cdots$&$0.41$&$0.29$\\
$110,426$& $\cdots$&$0.31$&$\mathbf{0.19}$&$0.26$&$\cdots$&$\cdots$&$0.27$\\
\hline
\end{tabular}

\textsuperscript{\emph{a}} As estimated from the graphs in Ref. \citenum{Schuett18};\\
\textsuperscript{\emph{b}} As estimated from the graphs in Ref. \citenum{Gub18}.
\end{sidewaystable*}

\begin{figure}
\includegraphics[width=\linewidth]{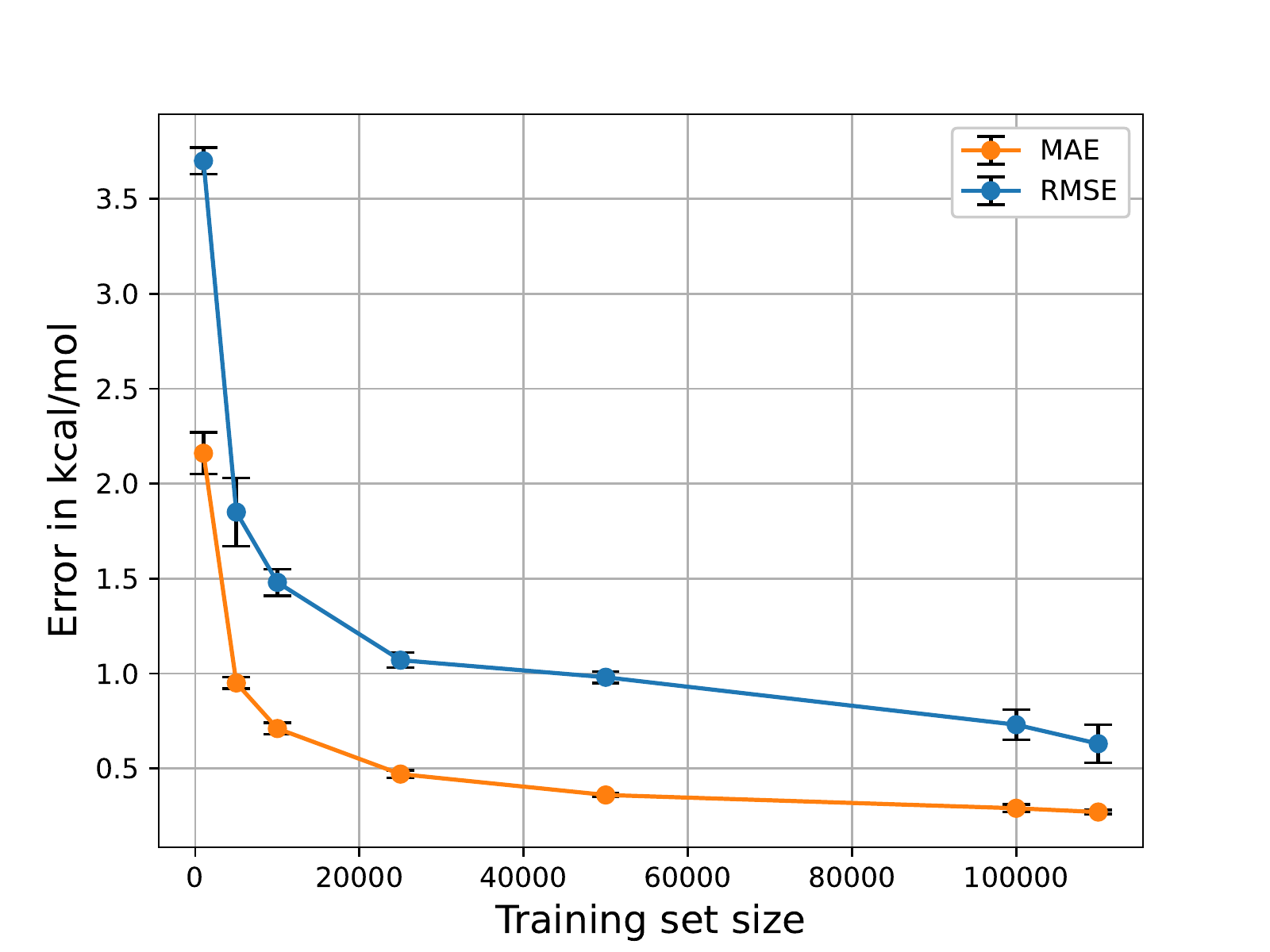}
\caption{Mean absolute error (MAE) and root mean square error (RMSE) in kcal/mol of the energy prediction on the QM9 data set depending on the number of structures in the training set. Results for all training set sizes are averaged over three and five independent randomly chosen training data sets, see text. The error bars indicate the standard deviation.}
\label{fig:fig_qm9}
\end{figure}

The learning curves of the model are shown in \figref{fig:fig_qm9}. They show the dependence of the MAE and the RMSE on the training set size.
For training set sizes of 1000, 5000, and 10,000 the results are obtained by averaging over five independent choices of the training set. For 25,000, 50,000, 100,000, and 110,426 structures only three independent choices of the training set are averaged. The GM-sNN trained on 110,426 reference energies predicts energies of the remaining structures with an MAE of $0.27$ kcal/mol and an RMSE of $0.63$ kcal/mol. The required accuracy of $1$ kcal/mol in the case of the MAE is reached already when training on 5000 reference structures. 

A comparison of the GM-sNN model to the various models published in the literature can be found in \Tabref{tab:table_qm9}. It can be seen that the performance of the GM-sNN model is comparable to all methods shown. However, one can see that the MTM$_{16-28}$ model~\cite{Gub18} and the model in Ref.~\citenum{Unke18} perform slightly better when training on 1000 and 5000 reference samples. The MTM$_{16-28}$ model employs geometric moments to construct rotationally invariant bases for linear regression. The model in Ref.~\citenum{Unke18} uses NNs as an ML method and the power spectrum of spherical harmonics as a structural descriptor. In both methods atomic species and environment are encoded simultaneously. With increasing number of training samples the GM-sNN model outperforms the MTM$_{16-28}$ model and the model in Ref.~\citenum{Unke18}. The GM-sNN model reaches an accuracy comparable to the message-passing models, e.g., SchNet~\cite{Schuett17, Schuett18}, HIP-NN~\cite{Lubb18}, and PhysNet~\cite{Unke19}. The message-passing models learn to construct invariant features from nuclear charges and interatomic distances in a data-driven manner. This approach was first introduced by the DTNN~\cite{Schuett17_2}.

We also investigated how well a model trained on small molecules transfers to larger systems. For this purpose the QM9 data set was divided into two subsets. The first subset contains molecules with up to $15$ atoms and has 24,978 structures in total. The other subset which is used for testing has molecules with more than $15$ atoms and has 105,853 structures in total. We used 22,978 structures of the first subset for training and another 2000 for validation. The errors on the test set of all 105,853 structures are averaged over three independent choices of the training set and are: $\text{MAE} = 1.01$ kcal/mol, $\text{RMSE} = 1.65$ kcal/mol. This demonstrates that the trained models can be transferred from small to large structures. However, the performance deteriorates compared with the randomly chosen structures, see \Tabref{tab:table_qm9}.

All models for the QM9 data set were trained on an NVIDIA Tesla V100-SXM2-32GB GPU. The training of 5000 epochs took from 1 hour (1000 structures) to 3 days (110,426 structures).

\subsection{\label{sec:sec3.2}MD17}

The MD17 data set~\cite{Chmiela17, Schuett17_2, Chmiela18} is a collection of structures, energies and atomic forces of eight small organic molecules obtained from ab-initio molecular dynamics (MD). For each molecule a large variety of conformations is covered. The data set varies in size from 150,000 to almost 1,000,000 conformations. It covers energy differences from $20$ to $48$ kcal/mol and force components ranging from $266$ to $570$ kcal/mol/\AA. The task of this experiment is to predict energies and forces for these molecules using various models.

We have chosen a cutoff radius of $R_\text{max} = 4.0$~{\AA}, since already $19$ of the $20$ possible neighboring atoms of the central atoms of the aspirin molecule (acetylsalicylic acid), the largest molecule of the MD17 data set, lie withing a sphere defined by this cutoff. 

\begin{figure}
\includegraphics[width=\linewidth]{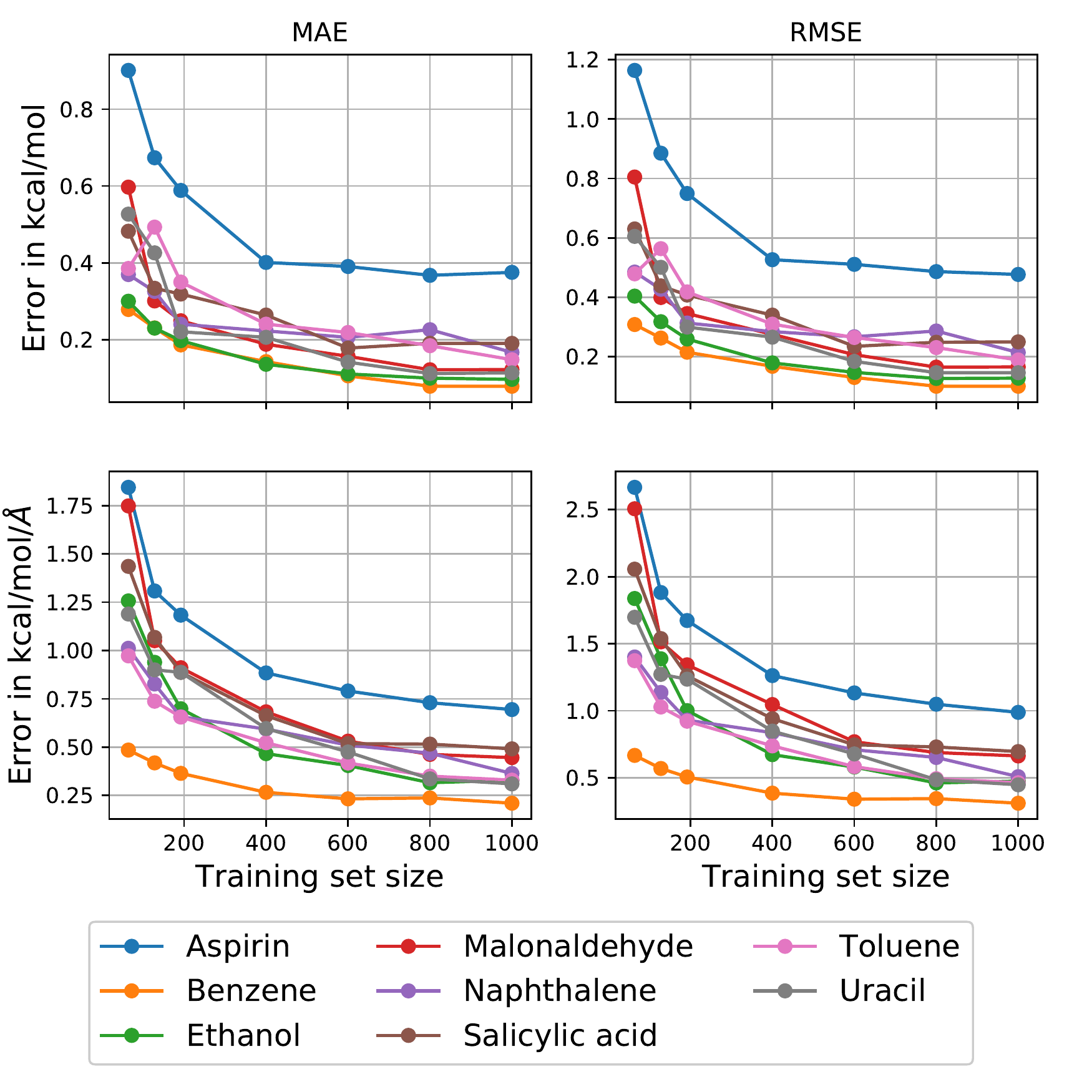}
\caption{Mean absolute error (MAE) and root mean square error (RMSE) in kcal/mol and kcal/mol/\AA{} of energy (top) and force (bottom) predictions, respectively, on the MD17 data set depending on the number of structures in the training data set. All results were obtained using the shallow architecture~GM-sNN.}
\label{fig:fig_md17}
\end{figure}

In a first test, we investigated the learning curves of the shallow GM-sNN model trained on structures from the MD17 data set. For this purpose we trained the model on $64$, $128$, $192$, $400$, $600$, $800$, and $1000$ randomly chosen samples. The respective learning curves for eight small organic molecules are presented in \figref{fig:fig_md17}.
From the figure it is noticeable that already $64$ samples are enough to achieve an accuracy of $1$~kcal/mol in energy. For most molecules at least 192 training samples are necessary to achieve an accuracy of 1 kcal/mol/\AA{} of the forces. Aspirin requires $400$ samples, but benzene requires only $64$ due to its rigid conformation. 

It may be noticed that the force learning curves look smoother than the energy learning curves. This is because a large weighting factor of $w_\text{F}=100$~\AA$^2${} was used in the loss function for the forces. Thus, most emphasis was given on the force training. In a typical example $\approx 99.7\%$ of the loss at the end of the training is caused by the forces. 
However, in all cases training could be continued which would lead to smaller force errors and to a higher impact of energies on the training. Further training would make the energy learning curves smoother.

\begin{sidewaystable*}[htbp]
\centering
\caption{\label{tab:table_md17}Mean absolute errors for energy and force prediction in kcal/mol and kcal/mol/\AA, respectively. The results are obtained by averaging over three independent choices of the training sets, their standard deviation is given in parentheses. The GM-NN models are trained on $1000$ and $50,000$ training samples. All models are trained on energies and forces with the exception of the GDML~\cite{Chmiela17} model, which is trained on forces only.}
\begin{tabular}{cccccl|clll}
\hline
& & \multicolumn{4}{c|}{$N = 1000$} &\multicolumn{4}{c}{$N = 50,000$} \\
\cline{3-6} \cline{7-10}
& & GDML~\cite{Chmiela17} & EANN~\cite{Zhang19} & SchNet~\cite{Schuett17} & GM-sNN & SchNet~\cite{Schuett17} & PhysNet~\cite{Unke19} & GM-sNN & GM-dNN\\
\hline
\multirow{2}{*}{\textbf{Benzene}} 	& energy & $\mathbf{0.07}$ & $\cdots$ & $0.08$ & $0.08~\left(0.008\right)$ & $\mathbf{0.07}$ & $\mathbf{0.07}~\left(0.002\right)$ & $\mathbf{0.07}~\left(0.003\right)$&$\mathbf{0.07}~\left(<0.001\right)$\\
 	& force  & $0.23$ & $\cdots$ & $0.31$ & $\mathbf{0.21}~\left(0.021\right)$ & $0.17$ & $0.15~\left(0.001\right)$ & $\mathbf{0.14}~\left(<0.001\right)$&$\mathbf{0.14}~\left(0.001\right)$\\

\multirow{2}{*}{\textbf{Toluene}} 	& energy & $0.12$ & $\mathbf{0.11}$ & $0.12$ & $0.15~\left(0.009\right)$ & $\mathbf{0.09}$ & $0.10~\left(0.004\right)$ & $0.10~\left(0.006\right)$&$\mathbf{0.09}~\left(0.003\right)$\\
	& force  & $\mathbf{0.24}$ & $0.38$ & $0.57$ & $0.34~\left(0.012\right)$ & $0.09$ & $\mathbf{0.03}~\left(0.002\right)$ & $0.10~\left(0.003\right)$&$0.06~\left(<0.001\right)$\\

\multirow{2}{*}{\textbf{Malonaldehyde}} 		& energy & $0.16$ & $0.14$ & $0.13$ & $\mathbf{0.12}~\left(0.012\right)$ & $0.08$ & $\mathbf{0.07}~\left(<0.001\right)$ & $\mathbf{0.07}~\left(0.003\right)$&$\mathbf{0.07}~\left(<0.001\right)$\\
 	& force  & $0.8$  & $0.62$ & $0.66$ & $\mathbf{0.45}~\left(0.014\right)$ & $0.08$ & $\mathbf{0.04}~\left(0.002\right)$ & $0.08~\left(0.006\right)$ &$0.05~\left(0.006\right)$\\

\multirow{2}{*}{\textbf{Salicylic acid}} 		& energy & $\mathbf{0.12}$ & $0.14$ & $0.20$ & $0.19~\left(0.020\right)$ & $\mathbf{0.10}$ & $0.11~\left(0.005\right)$ & $0.11~\left(0.002\right)$&$0.11~\left(0.002\right)$\\
	& force  & $\mathbf{0.28}$ & $0.51$ & $0.85$ & $0.49~\left(0.021\right)$ & $0.19$ & $\mathbf{0.04~\left(0.001\right)}$ & $0.14~\left(0.001\right)$&$0.08~\left(0.002\right)$\\
 
\multirow{2}{*}{\textbf{Aspirin}} 	& energy & $\mathbf{0.27}$ & $0.33$ & $0.37$ & $0.38~\left(0.015\right)$ & $\mathbf{0.12}$ & $\mathbf{0.12}~\left(0.005\right)$ & $0.19~\left(0.006\right)$&$0.13~\left(0.004\right)$\\
	& force  & $0.99$ & $0.99$ & $1.35$ & $\mathbf{0.69}~\left(0.025\right)$ & $0.33$ & $\mathbf{0.06}~\left(0.002\right)$ & $0.26~\left(0.009\right)$&$0.12~\left(0.008\right)$\\

\multirow{2}{*}{\textbf{Ethanol}} 	& energy & $0.15$ & $0.10$ & $\mathbf{0.08}$ & $0.10~\left(0.007\right)$ & $\mathbf{0.05}$ & $\mathbf{0.05}~\left(<0.001\right)$ & $\mathbf{0.05}~\left(<0.001\right)$&$\mathbf{0.05}~\left(0.002\right)$\\
 	& force  & $0.79$ & $0.47$ & $0.39$ & $\mathbf{0.33}~\left(0.017\right)$ & $0.05$ & $\mathbf{0.03}~\left(<0.001\right)$ & $0.06~\left(0.005\right)$&$0.04~\left(0.001\right)$\\
   
\multirow{2}{*}{\textbf{Uracil}} 	& energy & $\mathbf{0.11}$ & $\mathbf{0.11}$ & $0.14$ & $0.12~\left(0.008\right)$ & $\mathbf{0.10}$ & $\mathbf{0.10}~\left(0.001\right)$ & $\mathbf{0.10}~\left(<0.001\right)$&$\mathbf{0.10}~\left(0.001\right)$\\
		& force  & $\mathbf{0.24}$ & $0.35$ & $0.56$ & $0.33~\left(0.016\right)$ & $0.11$ & $\mathbf{0.03}~\left(<0.001\right)$ & $0.07~\left(0.005\right)$&$0.04~\left(<0.001\right)$\\
    
\multirow{2}{*}{\textbf{Naphthalene}} 		& energy & $\mathbf{0.12}$ & $\mathbf{0.12}$ & $0.16$ & $0.17~\left(0.011\right)$ & $\mathbf{0.11}$ & $0.12~\left(0.011\right)$ & $0.13~\left(0.017\right)$&$\mathbf{0.11}~\left(0.004\right)$\\
 	& force  & $\mathbf{0.23}$ & $0.27$ & $0.58$ & $0.36~\left(0.023\right)$ & $0.11$ & $\mathbf{0.04}~\left(0.001\right)$ & $0.13~\left(0.012\right)$&$0.08~\left(0.008\right)$\\
 \hline
\end{tabular}
\end{sidewaystable*}

A comparison of GM-NN models to several models recently published in the literature can be found in \Tabref{tab:table_md17}. The GM-sNN models were trained on $N=1000$ and $N=50,000$ samples, the GM-dNN models were trained on $N=50,000$ samples. The results of all models are averaged over three randomly chosen training sets. 
From \Tabref{tab:table_md17} we see that the GM-NN models yield an accuracy which is comparable with those of all well-established methods. The best training result is written in bold face. The shallow GM-sNN model outperforms the message-passing model SchNet when trained on $1000$ and 50,000 reference samples. The deep GM-dNN model reaches the accuracy of the PhysNet model. All mentioned message-passing models have more complicated mathematical forms and deeper NN architectures than our GM-NN models. Therefore, their capability of interpolation can potentially be better.

The GDML model is more accurate than our GM-sNN for the smaller molecules, although even there the difference is small, see \Tabref{tab:table_md17}.
Note that the GDML~\cite{Chmiela17} model was trained on forces only and, in general, scales badly with the number of reference structures due to its kernel nature. For small data sets and complex molecules, like aspirin, our GM-sNN model outperforms all presented methods in the force prediction. The force error on the aspirin data set is smaller by $0.3$~kcal/mol/\AA{} than the respective predictions of the GDML and EANN models, and smaller by $0.66$~kcal/mol/\AA{} than the SchNet predictions. The EANN model employs density-like descriptors and NNs as an ML method. The errors of GM-sNN in energy prediction could be improved training for more epochs, see the previous discussion. 

In addition to the models listed in \Tabref{tab:table_md17}, we can compare to sGDML~\cite{Chmiela18}, an extension of the GDML model that incorporates rigid space group symmetries and dynamic non-rigid symmetries, e.g. methyl group rotations.
The performance is similar. For example, the accuracy of the sGDML force prediction is $0.68$ kcal/mol/\AA{} for the aspirin data set, while GM-sNN results in $0.69$ kcal/mol/\AA{}.  The GM-sNN model needs fewer reference structures, less than $400$, to achieve an accuracy of $1$ kcal/mol/{\AA}, compared to the sGDML model, which needs about $600$ reference structures. Note that in this comparison it was assumed that the chosen training data is similarly correlated.

\begin{figure}
\includegraphics[width=\linewidth]{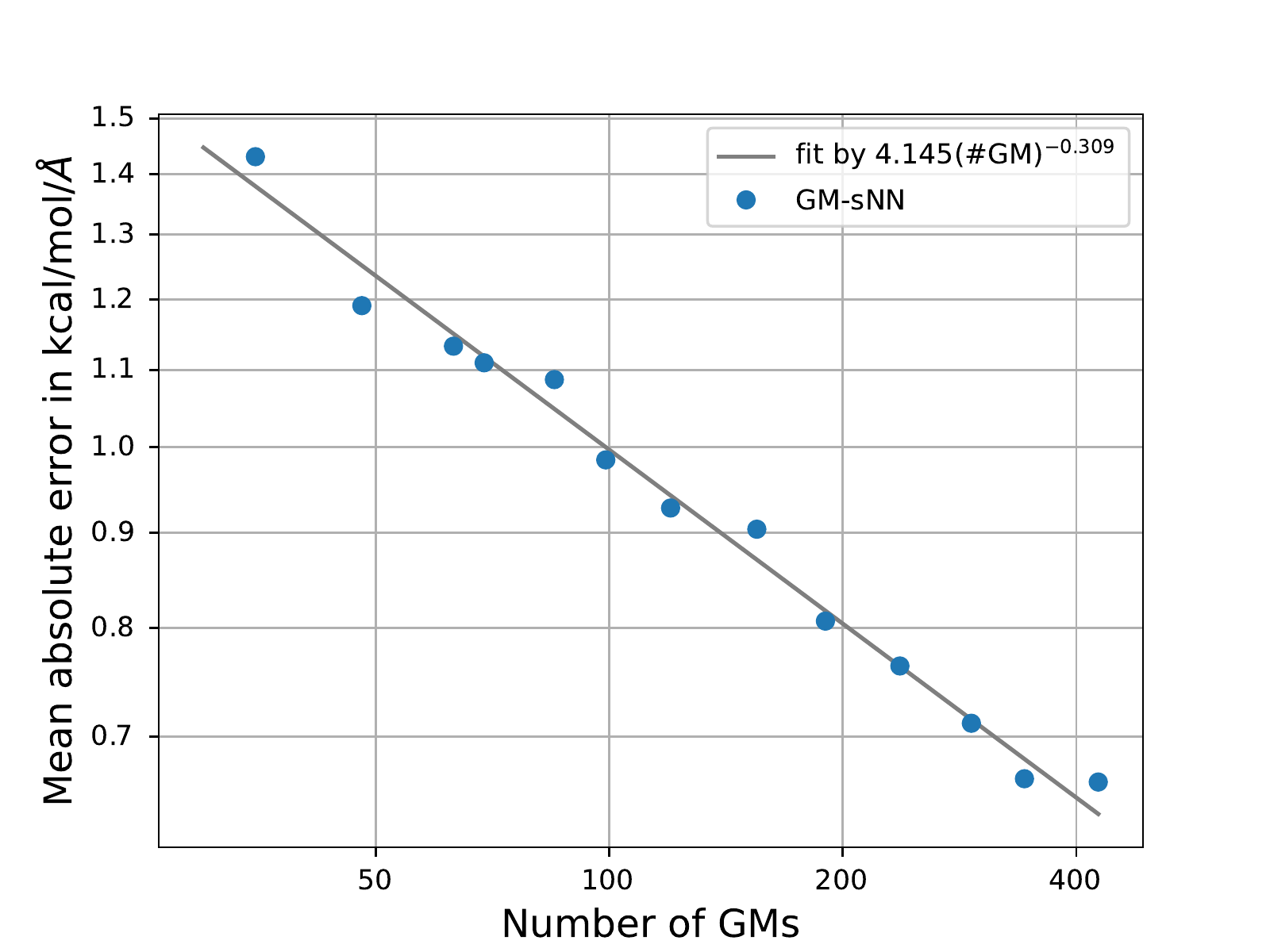}
\caption{Log--log plot of the algebraic decrease of the error in the predicted forces with an increasing number of Gaussian moments. All values are given on the test data of the aspirin data set for the GM-sNN model ($N=1000$).}
\label{fig:fig_convergence}
\end{figure}

We use the  MD17 data set  to test the dependence of the performance of the GM-sNN model on the size of our descriptor, the number of Gaussian moments (\#GM). \figref{fig:fig_convergence} shows that the force error is reduced algebraically with the increasing size of the descriptor. For aspirin and $N=1000$ we obtain an MAE of the forces of about $4.145\cdot (\text{\#GM})^{-0.309}$ kcal/mol/{\AA}. A similar algebraic convergence of the error in the energy prediction is illustrated in Fig.~S2 of the Supporting Information. To compare the performance with typical hand-crafted descriptors, atom-centered symmetry functions (ACSF)~\cite{Behler07, Behler10} were chosen. In Ref.~\citenum{Schuett19} it was shown that a typical Behler--Parrinello model with ACSFs as molecular descriptors is consistently outperformed by the SchNet model. For example, on the aspirin data set an MAE of $1.92$ kcal/mol/{\AA} in predicted forces was achieved using $51$ ACSF invariant scalars with 1000 training structures. For comparison, the GM-sNN model achieves an MAE of $1.43$ kcal/mol/{\AA} in predicted forces using only $35$ GM descriptors, and an MAE of $1.19$ kcal/mol/{\AA} with $48$ GM descriptors, see \figref{fig:fig_convergence}. This shows that the proposed descriptor outperforms ACSFs and captures all necessary information about the molecular structure as efficiently as message-passing architectures. Due to their particular mathematical form GMs achieve the desired flexibility and, thus, even GM-sNN outperforms the SchNet model in several tests, see \Tabref{tab:table_md17}.

All GM-NN models for the MD17 data set were trained on one NVIDIA Tesla V100-SXM2-32GB GPU each. The training of the GM-sNN model on $1000$ structures for $10,000$ epochs took $4$ hours, and the training on $50,000$ structures for $5000$ epochs was carried out during 2 days. The GM-dNN model required at most 2 days and 15 hours for the training.

\subsection{\label{sec:sec3.3} ISO17}

\begin{sidewaystable*}[htbp]
\centering
\caption{\label{tab:table_iso17}Mean absolute errors for energy and force prediction on the two variants of the ISO17 benchmark in kcal/mol and kcal/mol/\AA, respectively. The results are obtained by averaging over three independent choices of the training sets, their standard deviation is given in parentheses.}
\begin{tabular}{ccclll}
\hline
& & SchNet~\cite{Schuett17} & PhysNet~\cite{Unke19} & GM-sNN & GM-dNN \\
\hline
\multirow{2}{*}{\textbf{known molecules / unknown conformations}} 	& energy & $0.36$ & $\mathbf{0.10}~\left(<0.001\right)$ & $0.40~\left(0.020\right)$&$0.17~\left(0.003\right)$ \\
	& force  & $1.00$ & $\mathbf{0.12}~\left(0.002\right)$ & $0.65~\left(0.019\right)$&$0.28~\left(0.011\right)$ \\

\multirow{2}{*}{\textbf{unknown molecules / unknown conformations}} 	& energy & $2.40$ & $2.94~\left(0.260\right)$ & $\mathbf{1.97}~\left(0.414\right)$&$2.71~\left(0.640\right)$ \\
	& force  & $2.18$ & $\mathbf{1.38}~\left(0.060\right)$ & $1.66~\left(0.082\right)$&$1.96~\left(0.189\right)$ \\
\hline
\end{tabular}
\end{sidewaystable*}

The ISO17 data set~\cite{Rama14, Schuett17, Schuett17_2} contains short MD trajectories of $127$ isomers with the composition \ce{C7O2H10}, drawn randomly from the QM9 data set. For all molecules, energies and atomic forces are provided. Each trajectory samples $5000$ conformations. In total, the data set contains 635,000 structures. 

The experiment was arranged as follows. The data set was split into two subsets. The first subset contained the data of $\approx 80~\%$ of all molecules. From this subset 400,000 structures were taken randomly for training and another $4000$ structures were used for validation. The remaining 101,000 structures were used for testing the model. This test is referred to as ``known molecules / unknown conformations''. Then we applied the trained model to the remaining $\approx 20~\%$ of all molecules, those which the model had not seen before. This second test is referred to as ``unknown molecules / unknown conformations''. It allows to test the generalization capability of the GM-NN model.

The results of both tests obtained with the GM-sNN and GM-dNN models are compared to recent literature data in \Tabref{tab:table_iso17}. The results of the GM-NN models are obtained by averaging over three randomly chosen training sets. From the table it is noticeable that the GM-sNN model outperforms the SchNet model in 3 of the 4 tests. 
The shallow model also outperforms both message-passing models in the energy prediction for ``unknown molecules / unknown conformations''. The energy error is about $0.43$ kcal/mol lower than the SchNet prediction and $0.97$ kcal/mol lower than the PhysNet prediction. This shows that GM-sNN generalizes better than the models from the literature.

The deep GM-dNN model outperforms the shallow GM-sNN model and approaches the accuracy of PhysNet when applied to the ``known molecules / unknown conformations'' test. However, using the deep architecture deteriorates the performance on the ``unknown molecules / unknown conformations'' test. This indicates that the larger, more flexible network learns more details on the ``known molecules / unknown conformations'' test set on the expense of generalization capabilities, tested on the unknown molecules~\cite{neyshabur14}.
This example shows that a thorough choice of the network architecture is of crucial importance for the specific task for which the model is to be designed.

All GM-NN models were trained on one NVIDIA Tesla V100-SXM2-32GB GPU each for $5000$ training epochs. The training of the GM-sNN model took $\approx 7$ days, the training of the GM-dNN took $\approx$ 7 days and 6 hours. Note that the results of the PhysNet model were obtained after training for $\approx 1$ month~\cite{Unke19}.

\subsection{\label{sec:sec3.4}MD of Ethanol with Ab-Initio Accuracy}

\begin{figure}
\begin{center}
    \includegraphics[width=\linewidth]{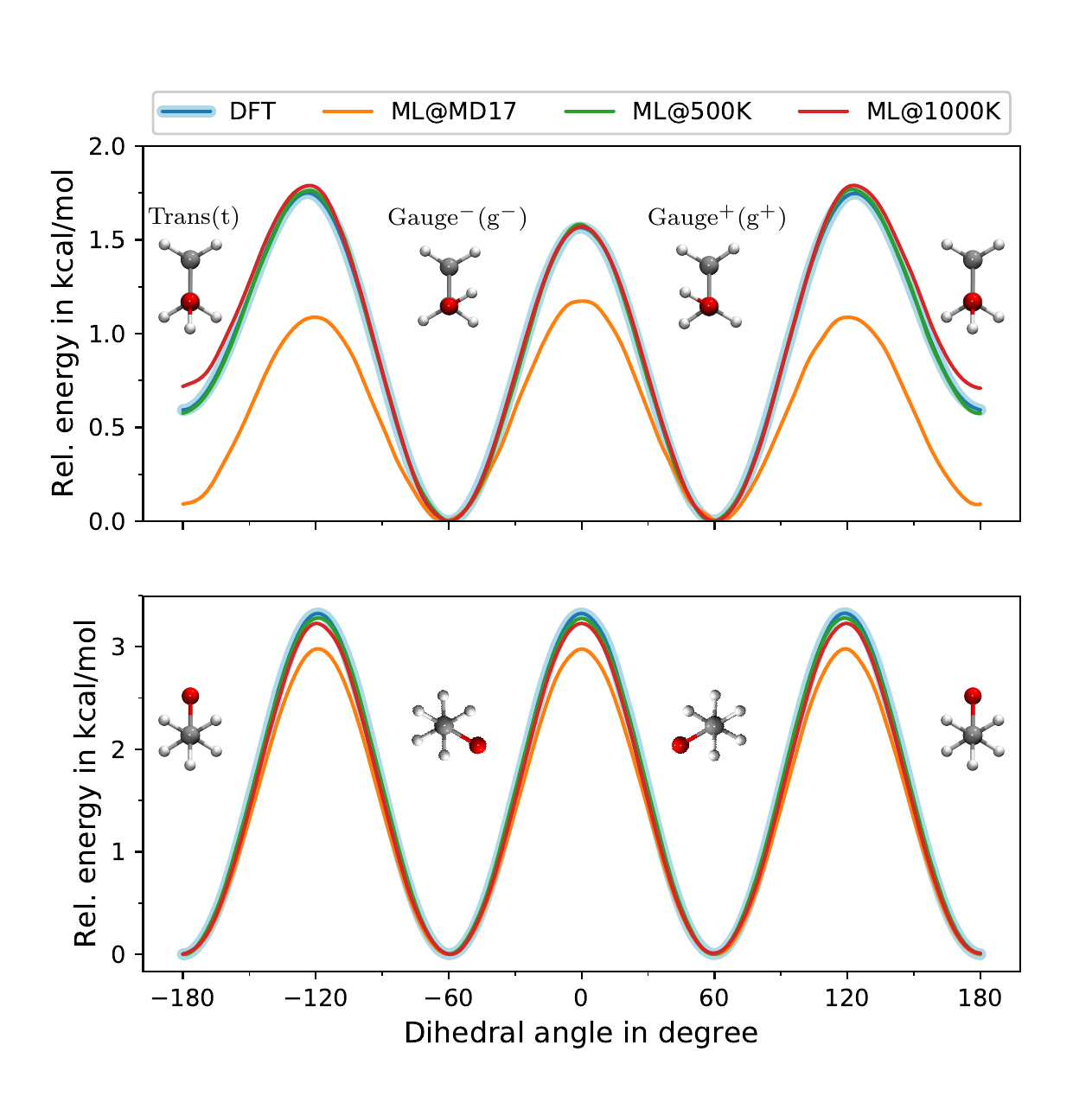}
\end{center}
\caption{Potential energy profile of the dihedral angle describing the rotation (top) of the OH-group around the C--O bond and (bottom) of the rotation of the CH$_3$-group around the C--C bond. The GM-sNN model was trained on the MD17 data set, ML@MD17, on the $500$~K data set, ML@500K, and on the $1000$~K data set, ML@1000K.}
\label{fig:fig_rotamere}
\end{figure}

The predictive power of the machine-learned potentials was tested on a simple organic molecule, namely ethanol. We calculated the energy profile for the ethanol rotamers, i.e., for the rotation of the OH-group around the C--O bond and the rotation of the \ce{CH3}-group around the C--C bond. A comparison of the predictions is made based on machine-learned potentials to the potential energy profile calculated at the PBE-D3(BJ)/6-31G* level of theory~\cite{PBE, Grimme-D3, Gimme-BJ, Rassolov98} using Turbomole 7.1~\cite{Turbomole} within ChemShell~\cite{Chemshell, She2003}
and is shown in \figref{fig:fig_rotamere}.

It is noticeable that the model trained on the MD17 data set for ethanol, ML@MD17, (we took the GM-sNN model trained on 50,000 structures, \secref{sec:sec3.2}) shows large deviations in the barrier heights. This is probably caused by the slightly different levels of theory: MD17 used PBE+vdW-TS (we were unable to find information on the basis used to create MD17~\cite{Chmiela17}). While the functionals are the same, the different treatment of dispersion and the different basis set in the reference may lead to the deviation of the energy profiles.

To ensure the reproducibility of the tests we generated two different data sets for ethanol on the same level of theory as for the respective DFT profile. The data sets were taken from ab-initio MDs at $500$ K and $1000$ K. In the following we describe the generation of the data sets. Firstly, we performed Born--Oppenheimer MD at $500$ K and $1000$ K in the $NVT$ ensemble using the Berendsen thermostat with GFN2-xTB~\cite{Grimme17, Bann19} as the underlying quantum mechanical method. The time step was set to $0.5$~fs and the dynamics was run for $50,000$ steps resulting in $25.0$~ps of dynamics. Every $10$ steps a geometry was taken from the dynamics and the energy, as well as atomic forces, were recalculated at the PBE-D3(BJ)/6-31G* level of theory. The MD was performed within ChemShell and for the refinement with DFT we used Turbomole 7.1 within ChemShell.
For each data set we obtained in total $5000$ structures. The additional data set at $1000$~K was created because the barrier for the rotation of the CH$_3$-group around the C--C bond is way higher than $500$~K. Both data sets can be found in a git-repository~\bibnote{\url{https://github.com/zaverkin/ethanol_datasets_git}}. 

The GM-sNN model was trained using $4000$ reference structures for $5000$ training epochs. Training of the model was performed on an NVIDIA Tesla V100-SXM2-32GB GPU and it took about $4.5$ hours for each data set. The remaining $1000$ structures were used for validation. We refer to the model trained on the $500$K data set as ML@500K and to the one trained on the $1000$K data set as ML@1000K. From \figref{fig:fig_rotamere} it can be seen that the model trained on the generated data sets fits the DFT profile well and all deviations are small. All barriers are given in \Tabref{tab:table_barriers}.

\begin{table}
\centering
\caption{Energetic barriers in kcal/mol predicted by machine learned potentials and calculated at the PBE-D3(BJ)/6-31G* level of theory.}
\label{tab:table_barriers}
\begin{tabular}{ccccc}
\hline
  &\multicolumn{3}{c}{OH}&\ce{CH3}\\
  \cline{2-4}
  &t$\rightarrow$g$^-$& g$^-$ $\rightarrow$t&g$^-\rightarrow$g$^+$&\\
\hline
PBE-D3(BJ)/6-31G* 	&$1.16$&$1.75$&$1.56$&$3.32$\\
ML@MD17 	&$1.00$&$1.09$&$1.17$&$2.98$\\
ML@500K 	&$1.19$&$1.76$&$1.58$&$3.28$\\
ML@1000K 	&$1.07$&$1.79$&$1.57$&$3.23$\\
\hline
\end{tabular}
\end{table}

To test the prediction of frequencies, even though only energies and forces were used for the training, we calculated the vibrational power spectrum of ethanol based on the ML@1000K model using the velocity-velocity autocorrelation function. In this formalism, the intensity of a transition is proportional to
\begin{equation}
\label{eq:eq_autocorr}
I \propto \left|\int \braket{v\left(t_0\right)v\left(t+t_0\right)}\exp\left(-i\omega t\right)\mathrm{d}t\right|^2.
\end{equation}

\begin{figure}
\centering
\includegraphics[width=\linewidth]{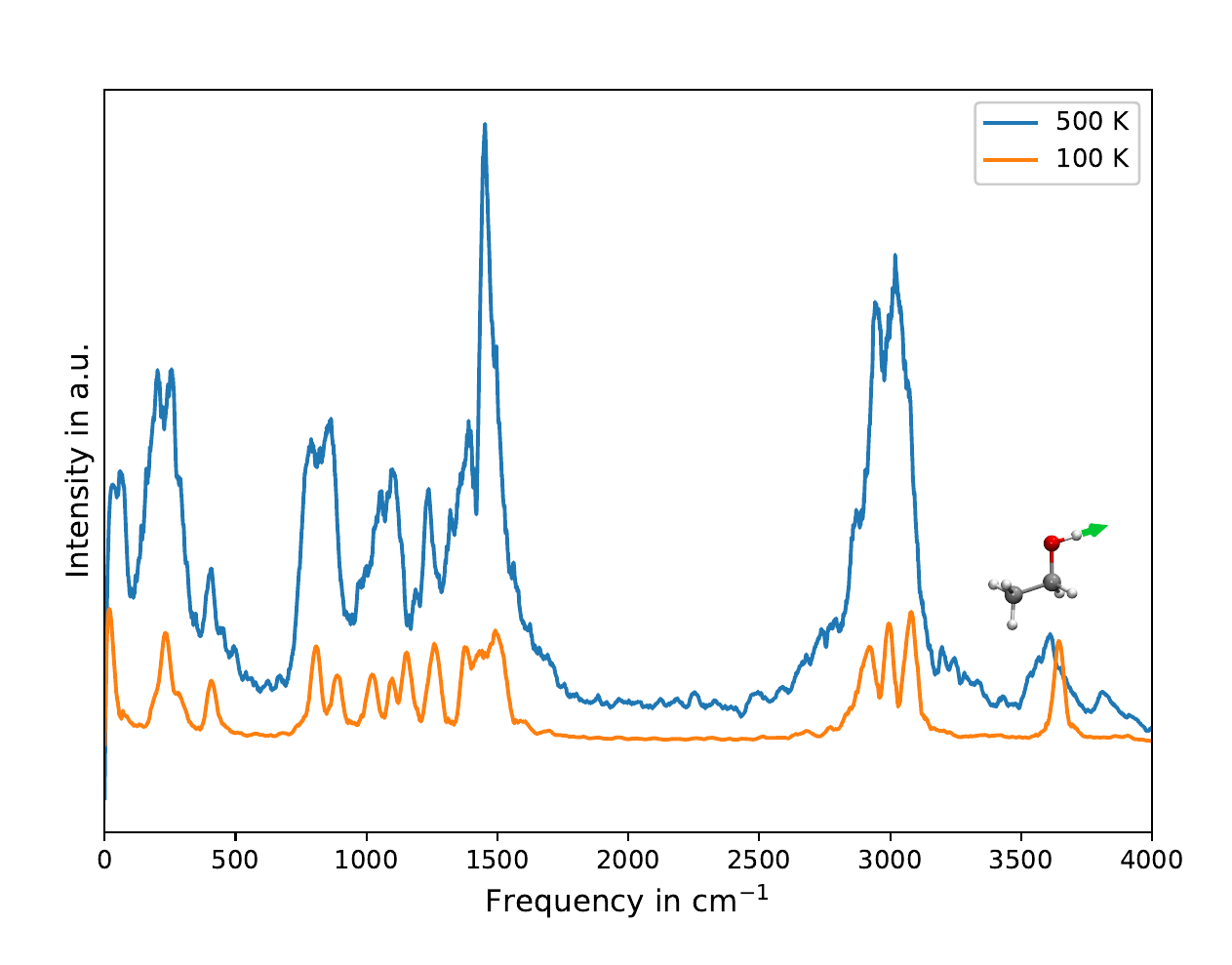}
\caption{Vibrational power spectrum of ethanol obtained via velocity-velocity autocorrelation function and the expression in \eqref{eq:eq_autocorr}. The velocities are obtained from MD at $500$~K and $100$~K using the GM-sNN model trained on the $1000$~K data set, see text for details.}
\label{fig:fig_spectra}
\end{figure}

Velocities for the calculation of the power spectrum were obtained by running MD trajectories on the ML@1000K model within ASE~\cite{ase17} using a Langevin thermostat at the temperatures of $500$~K and $100$~K. The time step was set to $0.5$~fs and the dynamics were run for $40$~ps. The first 1 ps was ignored. The final spectra obtained from MDs at $100$~K and $500$~K are shown in \figref{fig:fig_spectra}.

In \figref{fig:fig_spectra} one can, for example, find bands at $3611$~cm$^{-1}$ ($500$~K) and $3644$~cm$^{-1}$ ($100$~K) which correspond to the O--H stretching of alcohol. This is very similar to the corresponding harmonic frequency from DFT at the PBE-D3(BJ)/6-31G* level, $3638$~cm$^{-1}$. 
The experimental values for ethanol in the gas phase range from $3649$~cm$^{-1}$ to $3682$~cm$^{-1}$~\cite{NIST16}, which are also close to the values predicted using ML potentials. 

\section{\label{sec:sec4}Conclusions}

In the present work, we proposed Gaussian moments as a representation for molecular structures that incorporates global symmetries, i.e. the invariances with respect to rotation and translation of the entire system, and the invariance with respect to permutation of atoms of the same species. The particular advantage of constructing GMs is that the GM representation can be written in terms of pairwise distance vectors and tensor contractions. This allows for an efficient calculation of them on graphics processing units (GPUs). The representation can easily be extended by generating further rotationally invariant scalars from additional generating graphs. Thus, an even larger basis can be constructed if needed, at almost the same computational cost. 

We have demonstrated that the GM descriptor can be used as input for machine learning algorithms. In this work, we used feed-forward NNs as a machine learning method for the regression. We evaluated the GM-NN models on three different quantum-chemical benchmark data sets, which cover both chemical and conformational variability. Based on the performed tests we can argue that the GM-NN models show comparable or better accuracy with respect to the state-of-the-art machine learning models. The performance of GMs with only two hidden layers is similar to that of message-passing models, such as SchNet~\cite{Schuett17, Schuett18} and PhysNet~\cite{Unke19}, which have much deeper and mathematically more complicated NN architectures.

We have shown that a GM-model trained on small reference structures is able to generalize to larger structures. Additionally, it was shown that the respective GM descriptor is able to capture all necessary information about the molecular structure so that the machine learns as efficiently as respective models which include all possible symmetries explicitly.

In addition to the benchmark data sets, machine-learned potentials based on Gaussian moments were applied to predict rotamers and the vibrational power spectrum of the ethanol molecule. We have seen that the GM-NN potentials are capable of capturing differences between the gauge and trans conformations of ethanol and to capture vibrational frequencies even though they were trained on energies and forces only.

In summary, we have presented an approach for constructing a machine learning model based on tensor contractions, which fulfills physical constraints and is inspired by the molecular wave function. This model has been proven to be generally applicable to molecular systems and, therefore, can potentially be applied to large scale molecular simulations.

\section*{Acknowledgement}

The authors acknowledge financial support received in the form of a PhD scholarship from the Studienstiftung  des  Deutschen  Volkes (German National Academic Foundation).
We thank the Deutsche Forschungsgemeinschaft (DFG, German Research Foundation) for supporting this work by funding EXC 2075 - 390740016 under Germany's Excellence Strategy. We acknowledge the support by the Stuttgart Center for Simulation Science (SimTech) and the European Union's Horizon 2020 research and innovation programme (grant agreement No. 646717, TUNNELCHEM). We also like to acknowledge the support by the Institute for Parallel and Distributed Systems (IPVS) of the University of Stuttgart for providing computer time.

\section*{Supporting Information}

Additional data on the scalability and the computational cost, as well as a figure showing the decrease of the error in the energy prediction with the increase of the descriptor size are provided free of charge on the ACS Publications website.

\bibliography{references}

\end{document}